\documentclass[pdflatex,sn-aps,iicol]{sn-jnl}% American Physical Society (APS) Reference Style
\usepackage{graphicx}%
\usepackage{multirow}%
\usepackage{amsmath,amssymb,amsfonts}%
\usepackage{amsthm}%
\usepackage{mathrsfs}%
\usepackage[title]{appendix}%
\usepackage{xcolor}%
\usepackage{textcomp}%
\usepackage{manyfoot}%
\usepackage{booktabs}%
\usepackage{algorithm}%
\usepackage{algorithmicx}%
\usepackage{algpseudocode}%
\usepackage{listings}%
\theoremstyle{thmstyleone}%
\theoremstyle{thmstyletwo}%

\theoremstyle{thmstylethree}%

\begin{document}

\title{Angela and the electric dipole response -- giant and pygmy, hot and cold, isoscalar and isovector}

%%=============================================================%%
%% GivenName	-> \fnm{Joergen W.}
%% Particle	-> \spfx{van der} -> surname prefix
%% FamilyName	-> \sur{Ploeg}
%% Suffix	-> \sfx{IV}
%% \author*[1,2]{\fnm{Joergen W.} \spfx{van der} \sur{Ploeg} 
%%  \sfx{IV}}\email{iauthor@gmail.com}
%%=============================================================%%

\author[1,2]{\sur{Peter von Neumann-Cosel}}\email{vnc@ikp.tu-darmstadt.de}
\affil[1]{Institut f\"ur Kernphysik, Technische Universit\"at Darmstadt, 64289 Darmstadt, Germany}
%\orgdiv{Department}, \orgname{Organization}, \orgaddress{\street{Street}, \city{City}, \postcode{100190}, \state{State}, \country{Country}}}
\affil[2]{Norwegian Nuclear Research Center and Department of Physics, University of Oslo, N-0316 Oslo, Norway}
%\orcidlink{0000-0002-0256-5940}
%{\orgdiv{Department}, \orgname{Organization}, \orgaddress{\street{Street}, \city{City}, \postcode{10587}, \state{State}, \country{Country}}}

%%==================================%%
%% Sample for unstructured abstract %%
%%==================================%%

\abstract{
The impact of Angela Bracco's work on the electric dipole response of nuclei is discussed using three examples of current nuclear structure problems: disentangling different contributions to the decay width of the giant dipole resonance, the equivalence of photoabsorption and emission and the nature of the pygmy dipole resonance. 
}


%%\pacs[JEL Classification]{D8, H51}

%%\pacs[MSC Classification]{35A01, 65L10, 65L12, 65L20, 65L70}

\maketitle

\section{Introduction}
\label{sec:1}

Angela Bracco's contributions to nuclear physics cover a wide range of topics from few-body problems \cite{bracco83,bracco84}, nuclear structure phenomena like shape isomerism \cite{leoni17}, superdeformed bands \cite{lagergren01,jensen02}, low-energy structure in exotic nuclei \cite{zanon23,paxman25} to the damping of giant resonances \cite{bracco88,bracco89,bortignon91}.
However, there are two topics accompanying most of her research career: the advancement of $\gamma$ spectroscopy \cite{bracco02,bracco21} and the electric dipole response of nuclei \cite{bracco19}.
Of course, the two are related by the dominance of $E1$ radiation in the $\gamma$ decay of nuclear excited states.
Her work covers the whole range of phenomena like the pygmy dipole resonance (PDR) at low and the giant dipole resonance (GDR) at high excitation energies as well as studies of ground state strength and at finite temperatures using isoscalar and isovector probes.

In this contribution to the "Topical issue dedicated to Angela Bracco on the occasion of her farewell from Milano University", I try to reflect on how Angela's work has impacted on my own research. 
In fact, the influences are manyfold but I will constrain myself to three themes: (i) the damping of giant resonances exemplified by examples of studies of the isovector giant dipole resonance (IVGDR), (ii) the role of the Brink-Axel hypothesis for a statistical description of $\gamma$ absorption and decay and (iii) the nature of the low-energy structure observed in the $E1$ response of heavy nuclei commonly termed pygmy dipole resonance (PDR).

\section{Fine structure and the damping of giant resonances}
\label{sec:2}

\begin{figure}
\centering
\includegraphics[width=\columnwidth]{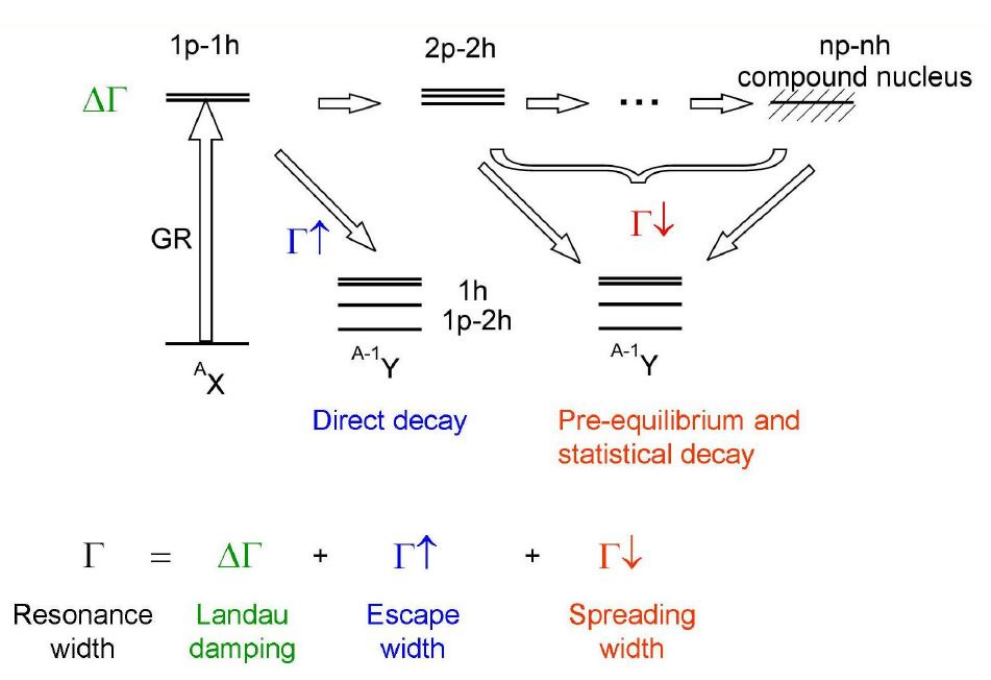}
\caption{Contributions to the decay width of a giant resonance.
For details see text. 
Reprinted from Ref.~\cite{vonneumanncosel19}.}
\label{fig:1}
\end{figure}

Interacting fermion systems with finite particle number like the nucleus form collective modes involving most of the particles, commonly called giant resonances \cite{harakeh01}. 
They appear at high excitation energies well above the particle thresholds and show widths of several MeV.
Much experimental work has gone into trying to establish their global features like centroid energies and widths. 
While the former are reasonably well understood in terms of simple mass dependence, describing the widths remains a challenging theoretical problem.
The total width $\Gamma$ of the resonance is understood to arise from  three mechanisms illustrated in Fig.~\ref{fig:1}: 
fragmentation of the elementary one particle-one hole ($1p$-$1h$) excitations (Landau damping $\Delta \Gamma$),
direct particle emission from $1p$-$1h$ configurations leading to an escape width $\Gamma^\uparrow$, and the mixing of $1p$-$1h$ configurations into more complicated two-particle two-hole ($2p$-$2h$) and finally to many particle-many hole ($np$-$nh$) states giving rise to a spreading width $\Gamma^\downarrow$
\begin{equation}
\label{eqwidth}
\Gamma = \Delta\Gamma + \Gamma^\uparrow + \Gamma^\downarrow.
\end{equation}

One important method to differentiate the role of these components is the coincident measurement of particle \cite{harakeh01} or $\gamma$ \cite{beene90} decay, where contributions of the escape width can be identified by the population of $1h$ and $1p$-$2h$ states in the daughter nucleus and the spreading width contribution can be estimated by comparison with statistical model calculations (see, for example, Refs.~\cite{bracco88,bolme88,diesener94,strauch00,hunyadi03}).
The scheme outlined implies a fragmentation of the giant resonance strength in a hierarchical manner leading to the picture of doorway states.
Angela Bracco and her coworkers Pier Francesco Bortignon and Ricardo Broglia at the University of Milano have made important contributions to a solution of this problem \cite{bortignon98}.

\begin{figure}
\centering
\includegraphics[width=\columnwidth]{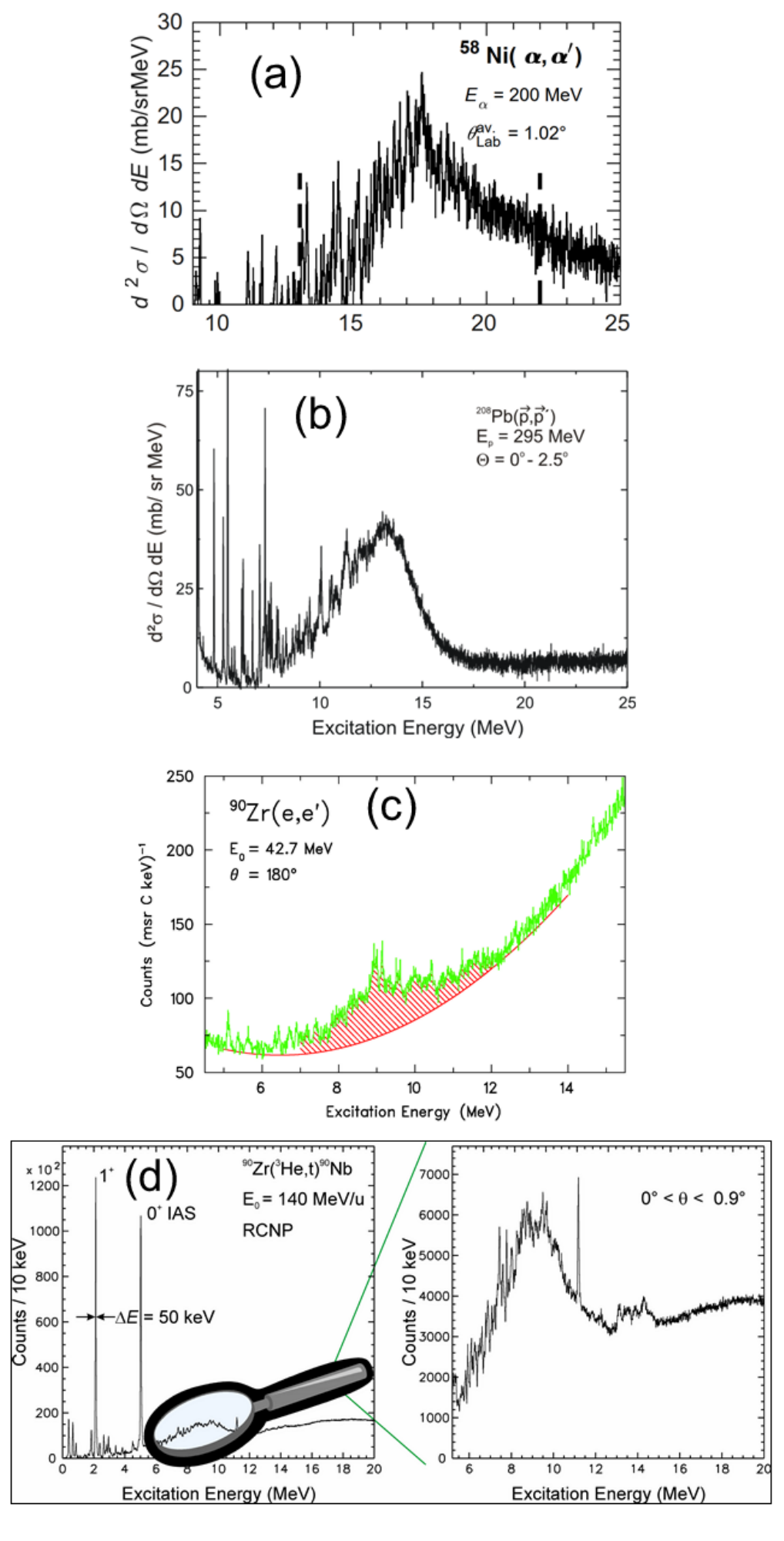}
\caption{Examples of the fine structure observed in high-resolution experiments for different types of giant resonances. 
(a) GMR in $^{58}$Ni. 
Reprinted from Ref.~\cite{bahini23}.
(b) GDR in $^{208}$Pb.
Reprinted from Ref.~\cite{tamii11}.
(c) $M2$ resonance in $^{90}$Zr.
Reprinted from Ref.~\cite{vonneumanncosel99}.
(d) GT resonance in $^{90}$Nb.
Reprinted from Ref.~\cite{kalmykov06}.
}
\label{fig:2}
\end{figure}

Experimentally, it is expected that the coupling scheme leads to fine structure of the giant resonances.
While studies of the gross features of giant resonances typically have limited energy resolution, in the last 20 years a series of experiments has been  performed trying to establish and understand the fine structure phenomenon \cite{vonneumanncosel19a}.
To summarize, fine structure was observed for the isoscalar giant monopole(GMR)  \cite{olorunfunmi22,bahini24}, dipole (GDR) \cite{poltoratska14,jingo18,donaldson21} and quadrupole (GQR) \cite{shevchenko04,shevchenko09} resonances as well as for magnetic modes like the GT \cite{kalmykov06} and $M2$ \cite{vonneumanncosel99} resonances, cf.\ Fig.~\ref{fig:2} for some representative examples. 
It appears across the nuclear chart and is even seen in heavy deformed nuclei \cite{donaldson21,kureba18}, where level densities are extremely high. 

A variety of methods based on a local scaling dimension approach \cite{aiba99,aiba11}, an entropy index method \cite{lacroix99,lacroix00}, Fourier analysis \cite{heiss11} and the use of wavelet techniques \cite{shevchenko04} was proposed to extract quantitative information on the fine structure.   
A comprehensive comparison of the methods showed that wavelet analysis is best adapted to the problem \cite{shevchenko08}.

\begin{figure}
\centering
\includegraphics[width=\columnwidth]{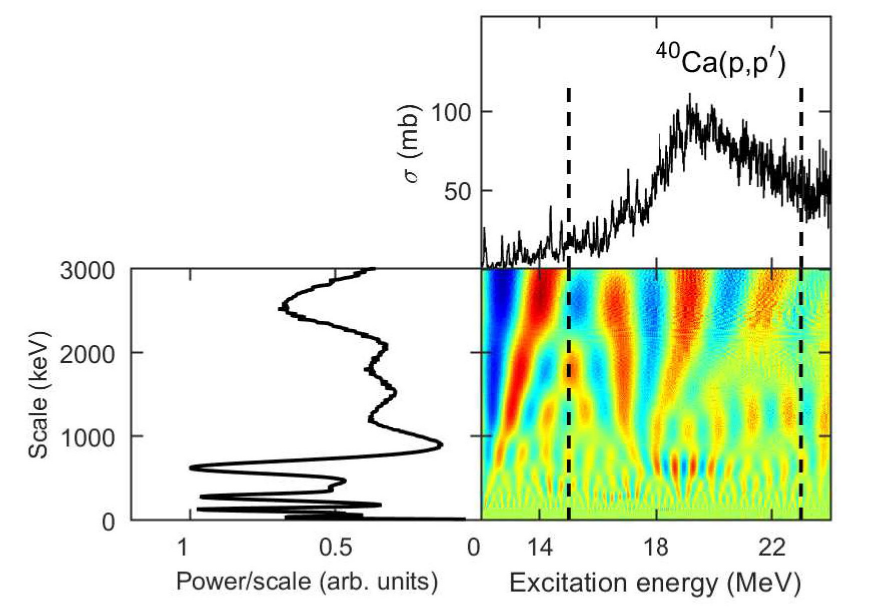}
\caption{Wavelet analysis of the photoabsorption cross sections deduced from the $^{40}$Ca($p,p^\prime$) reaction.
For details see text.
Reprinted from Ref.~\cite{carter22}.}
\label{fig:3}
\end{figure}

The kind of information that can be extracted from the wavelet analysis is illustrated by a recent study of the GDR fine structure in $^{40,48}$Ca using inelastic proton scattering measured at extreme forward angles \cite{carter22}.
At beam energies of several hundred MeV this reaction selectively excites $E1$ strength by relativistic Coulomb excitation \cite{vonneumanncosel19}.
The upper right part of Fig.~\ref{fig:3} shows a spectrum of $^{40}$Ca measured at a laboratory scattering angle of about $1^\circ$.  
The wavelet analysis proceeds via the calculation of a wavelet coefficient $C$ from the measured cross sections $\sigma(E)$  
\begin{equation}
C(\delta E,E_x)=\frac{1}{\sqrt{\delta E}} \int \sigma(E) \Psi^*\left(\frac{E_x-E}{\delta E}\right) dE,
\label{eq:wc}
\end{equation}
where $E_x$ is the excitation energy, $\delta E$ the wavelet scale, and $\Psi$ the wavelet function. 
For studies of nuclear giant resonances, the complex Morlet wavelet 
\begin{equation}
\Psi(x)=\pi^{-1/4}\,e^{ik_0x}\,e^{-x^2/2},
\label{eq:wf}
\end{equation}
with $k_0=5$ is employed.
The continuous wavelet transform (CWT) is used, where the fit procedure can be adjusted freely to the required precision providing optimum balance between resolution of excitation energy and energy scale.
Further details can be found in Ref.~\cite{shevchenko08}. 

By way of example, the real part of the wavelet coefficients is displayed in the bottom-right part of Fig.~\ref{fig:3}.
Large positive values are indicated by shades of red going 
down to close to zero in yellow with negative coefficients in shades of blue.
At certain wavelet scale values maxima of the absolute values are observed across the excitation region of the GDR.
It is convenient to project the two-dimensional distribution on the scale axis to visualize them.
The resulting power spectrum
\begin{equation}
P_w(\delta E)=\frac1N \sum_{i=i_1}^{i_2}|C_i(\delta E)C^*_i(\delta E)|,
\label{eq:power}
\end{equation}
where $i_1$ and $i_2$ indicate the boundaries of the region of interest indicated by the vertical dashed lines, is shown in the bottom left part of Fig.~\ref{fig:3}.
Peaks of strength in this power spectrum are associated with characteristic scales of the fine structure in the region of the GDR. 

\begin{figure}
\centering
\includegraphics[width=\columnwidth]{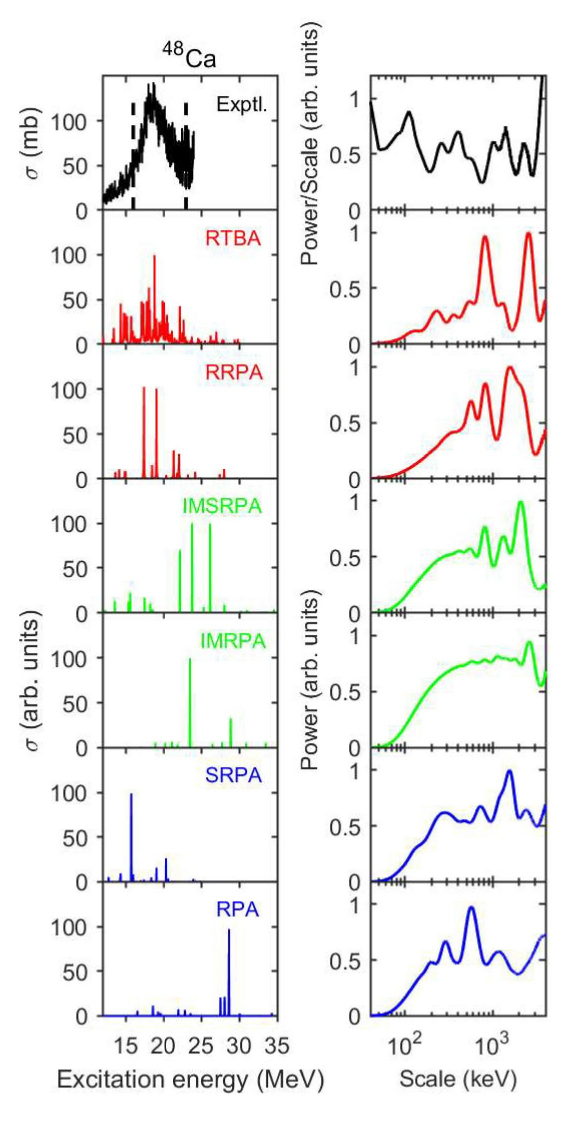}
\caption{Experimental and theoretical photoabsorption cross sections of $^{48}$Ca and corresponding power spectra of the wavelet analysis.
Reprinted from Ref.~\cite{carter22}, where details on the model calculations can be found.
}
\label{fig:4}
\end{figure}

While Fig.~\ref{fig:3} illustrates how quantitative information on the fine structure can be extracted, its interpretation requires a comparison to model calculations.
The same type of wavelet analysis can be performed for theoretical strength functions.
Figure~\ref{fig:4} shows  photoabsorption cross sections of $^{48}$Ca and the wavelet power spectra from experiment and various theoretical results based on RPA and beyond RPA calculations.
The doubly magic nuclei $^{40}$Ca and $^{48}$Ca are particularly suited because they permit second-RPA (SRPA) calculations including $2p-2h$ degrees of freedom on top of the $1p-1h$ model space in RPA. 
While the wavelet power spectra at RPA level indicate the possible relevance of Landau fragmentation, the change when including $2p-2h$ states provides a measure of the importance of the spreading width. 
A detailed discussion is provided in Ref.~\cite{carter22}.

The comparison of damping mechanisms in GMR, GDR and GQR reveals siginifcant differences.
While fine structure of the GQR is dominated by coupling to $2p-2h$ states \cite{shevchenko04,shevchenko09}, or more specifically to low-lying vibrations (see, however, Ref.~\cite{usman11}), Landau fragmentation dominates for the GDR \cite{vonneumanncosel19a}.
Recent study of the GMR lie somewhere in between \cite{bahini24}.
For future work, it would be important to explicitly treat the continuum known to be important in lighter nuclei on top of $2p-2h$ degrees of freedom. 

\section{E1 response at finite temperature and the Brink-Axel hypothesis}
\label{sec:3}

\begin{figure}[b]
\centering
\includegraphics[width=\columnwidth]{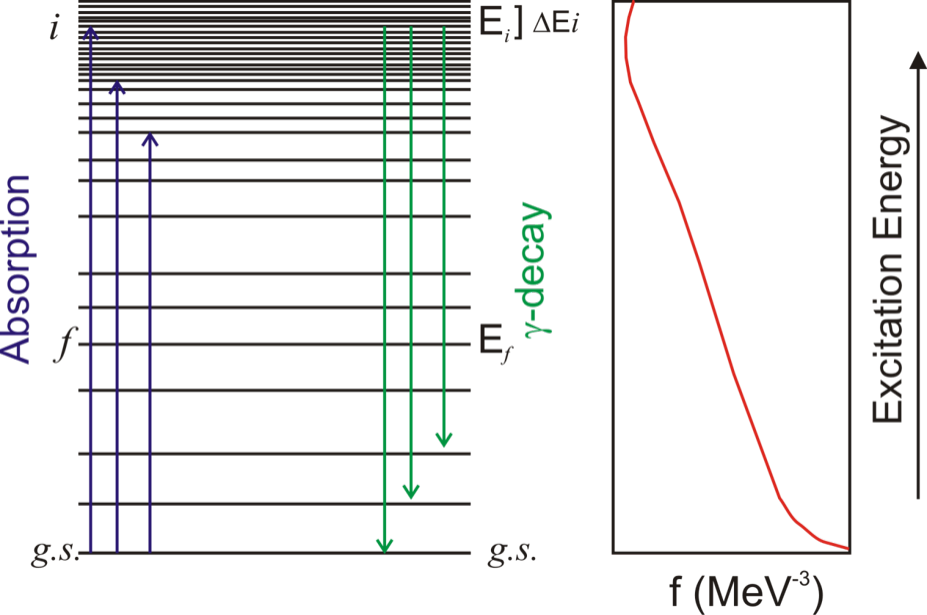}
\caption{Left: Ground- and excited-states photoabsorption andemission.
Right: Scheme of the GSF.}
\label{fig:5}
\end{figure}

Understanding nuclear structure requires systematic studies of the nuclear response as a function of key parameters like mass, spin or excitation energy.
The latter is of particular importance for the description of nuclear reactions in stellar environments, where temperatures can be sufficiently high that in equilibrium significant parts of the nuclei are in excited states. 
This immediately raises the question whether the strength functions change as a function of temperature.
Theoretical work based on shell model \cite{johnson15,sieja23} or QRPA \cite{hung17,wibovo22} predict this is generally the case for collective modes, except maybe for the IVGDR.
In any case, finite-temperature QRPA predicts strong modifications of the $E1$ response at low $\gamma$ energies, see e.g. Ref.~\cite{kaur25} and references therein.

\begin{figure*}
\centering
\includegraphics[width=\textwidth]{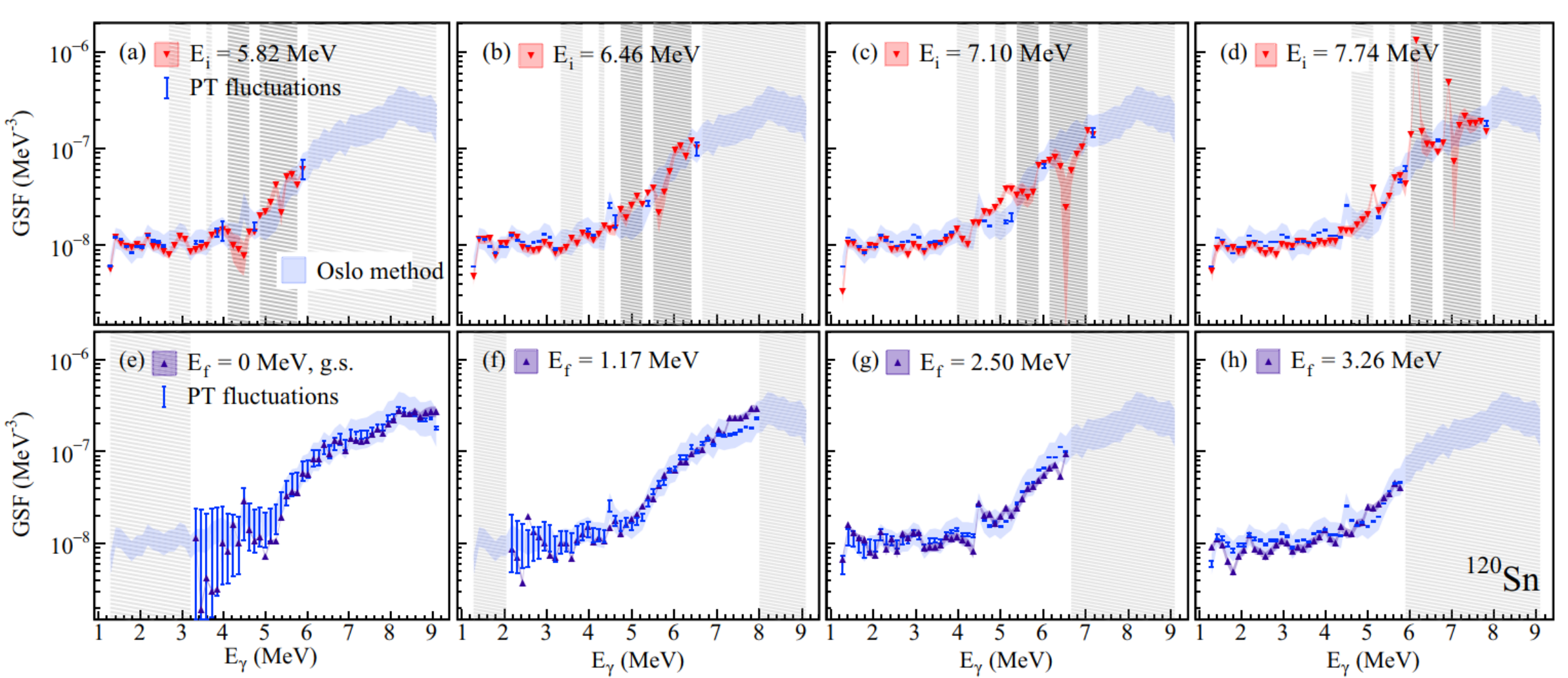}
\caption{GSF of $^{120}$Sn with gate on several initial (top row, red triangles) and final (bottom row, blue triangles) energies compared with the Oslo method strength (blue shaded band). 
For each strength the statistical error band is shown together with the error due to Porter-Thomas ﬂuctuations. 
Dark gray regions correspond to the areas of expected inﬁnite Porter-Thomas ﬂuctuations, light gray area marks energies for which the ﬂuctuations of the strength were not determined.
Reprinted from Ref.~\cite{markova22}.}
\label{fig:6}
\end{figure*}

For the latter, the Brink-Axel (BA) hypothesis \cite{brink55,axel62} was put forward that the electric dipole strength distribution is independent on whether it is built on the ground state or an excited state. 
The concept is illustrated in the left-hand side of Fig.~\ref{fig:5}. 
A typical $\gamma$ strength function (GSF) is shown on the right-hand side.
It is dominated by the GDR, but for many astrophysical applications only the part below the neutron threshold is relevant. 
There, a resonance-like structure is observed in neutron-rich nuclei called pygmy dipole resonance (PDR) discussed in more detail in Sec.~\ref{sec:4}.
Nucleosynthesis reaction network calculations are typically performed within a statistical approach, and the BA hypothesis dramatically simplifies the problem for reactions including $\gamma$ absorption or emission.
However, there is no real justification and its applicability is a topic of controversial discussion.  

The observation of the GDR in the $\gamma$ decay of highly excited compound nuclei permits a study of the temperature dependence, and Angela Bracco has been a key figure in this research \cite{bracco89,bortignon91,broglia92,bracco95}.
At temperatures of several MeV, the GDR exhibits shape changes violating the BA hypothesis, but for $T \leq 1$ MeV an approximate constancy is observed.
For the GSF below neutron threshold most important in astrophysical applications, conflicting results have been reported claiming confirmation \cite{guttormsen16,martin17,crespocampo18,scholz20} or violation \cite{angell12,isaak13,netterdon15,isaak19} of the BA hypothesis.  

This problem has been recently adressed  in a study of the $E1$ response in the Sn isotope chain combining ground-state photoabsorption measured with relativistic Coulomb excitation \cite{bassauer20a} with quasicontinuum $\gamma$ decay data \cite{markova21} analyzed with the Oslo method \cite{schiller00}.
The latter results permit a test of the independence of the GSF from initial and final energy demonstrated in Fig.~\ref{fig:6}  for the case of $^{120}$Sn \cite{markova22}.
The upper  and lower rows show the GSFs extracted for specific excitation energy (red circles)  or final energy (blue circles) windows. 
Clearly, they are all compatible within uncertainties with the result averaged over the full accessible energy range (light blue band) as expected from the BA hypothesis.

\begin{figure}[b]
\centering
\includegraphics[width=\columnwidth]{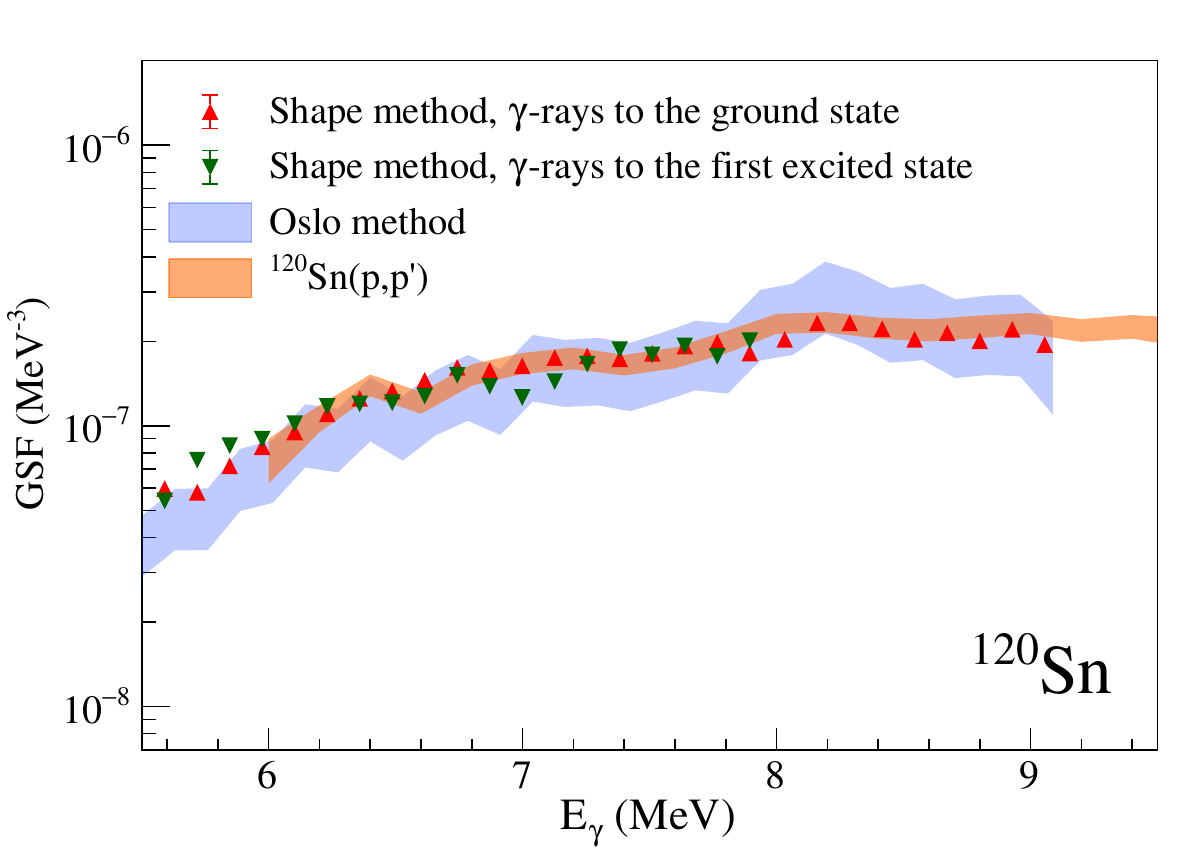}
\caption{ Comparison of the GSF in $^{120}$Sn extracted with the
Oslo method (blue band), from selective decay to the ground state
and the first excited $2^+$ state utilizing the shape method \cite{wiedeking21} (red and green triangles) and from $(p,p^\prime)$ scattering at forward angles (orange band).
Reprinted from Ref.~\cite{markova21}.}
\label{fig:7}
\end{figure}

The direct comparison between photoabsorption and -decay in $^{120}$Sn is shown in Fig.~\ref{fig:7} as orange and blue bands, respectively.
In the energy region covered by both experiments, the GSFs agree within error bands. 
The shape method \cite{wiedeking21}, where the GSF is derived from decay to separated low-lying states, provides yet another independent test.
In the present case, decay to the ground state and first excited $2^+$ state were experimentally resolved.  
The shape method does not yield absolute values, but after normalization to the $(p,p^\prime)$ data, the resulting GSFs shown as red and green triangles are in very good agreement. 

These comprehensive results -- confirmed in other Sn isotopes -- demonstrate that the BA hypothesis holds for low-energy $E1$ strength, at least within typical experimental uncertainties.
However, it is not clear yet to what extent they can be generalized.
Future work has to explore the limits systematically with respect to mass, excitation energy and level density.
Limits for the latter are e.g.\ indicated by deviations of the shape method at energies below 5 MeV. 

\section{Nature of the PDR}
\label{sec:4}

A current topic of low-energy nuclear physics is the observation of a local enhancement of the electric dipole response in heavy nuclei with neutron excess, commonly called PDR \cite{savran13}.
Besides understanding the underlying structure, it impacts on $(n,\gamma$) cross sections in various astrophysical processes contributing to the nucleosynthesis of heavy elements \cite{wiedeking24}.
Angela Bracco has made decisive contributions \cite{bracco19} to its understanding, in particular to measurements of the isovector strength in exotic neutron-rich nuclei \cite{wieland09,wieland18}, its isoscalar response \cite{crespi14,pellegri14,crespi21} and the possible connection to the equation of state of neutron-rich matter \cite{carbone10}. 

\begin{figure}
\centering
\includegraphics[width=\columnwidth]{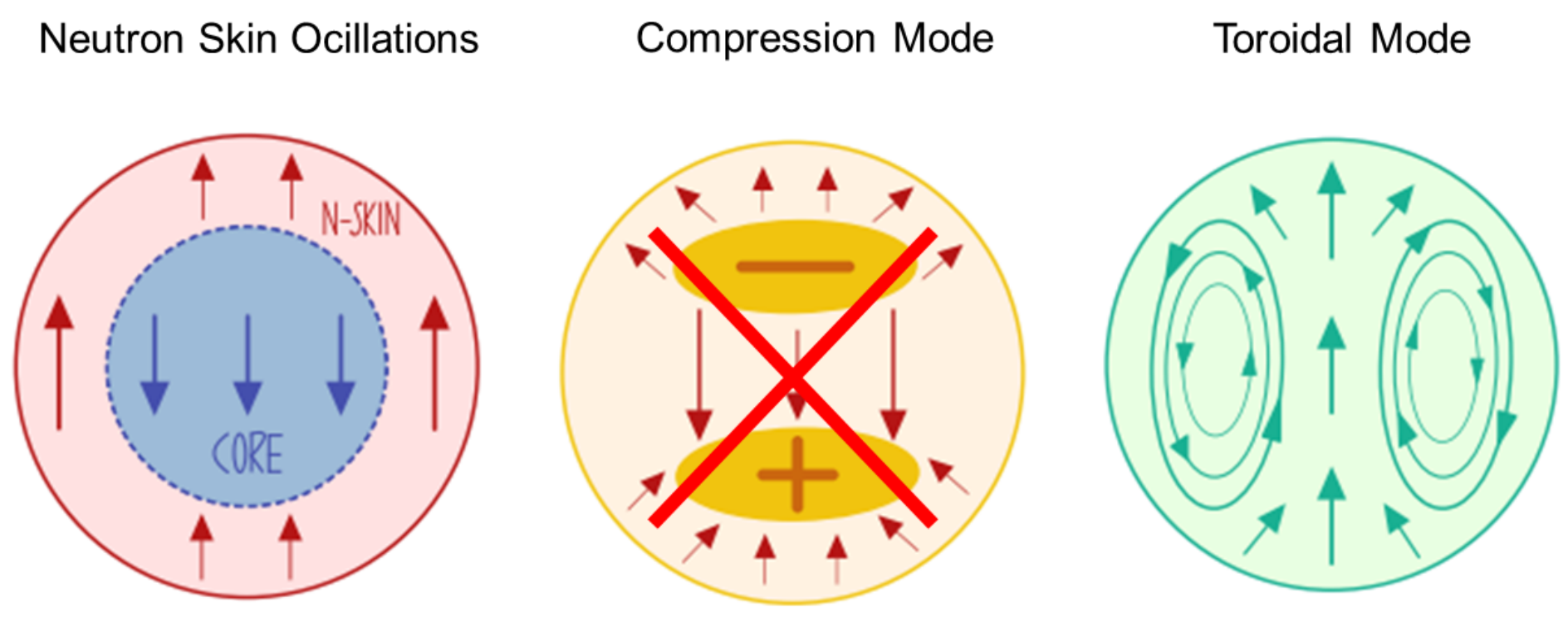}
\caption{Schematic velocity distributions of different interpretations of the PDR as neutron skin oscillation, compressional mode, or toroidal mode. 
Figure adapted from Ref.~\cite{lanza23}.}
\label{fig:8}
\end{figure}

The nature of the PDR is a subject of long-standing discussion \cite{lanza23}.
Three possible interpretations are illustrated in Fig.~\ref{fig:8} by their characteristic current distributions.
It may be seen as oscillation of the neutron skin relative to an isospin-saturated core \cite{paar07}, as low-energy branch of the dipole compression mode \cite{rocamaza18}, or as a toroidal mode \cite{repko13,repko19}.
An interpretation as low-energy compression mode can most likely be excluded because experimental isoscalar strengths are so large \cite{poelhekken92} that it would significantly alter the compressibility extracted from the isoscalar GDR to disagree with the value derived from the GMR \cite{rocamaza18}.

The claim of the PDR representing a neutron skin oscillation is mainly based on the peculiar form of the radial transition density in density functional models, which shows an approximately isoscalar behavior in the interior and a pronounced peak of the neutron part at the nuclear surface.
If true, its isovector strength and evolution with neutron excess could provide information on the neutron skin thickness and parameters of the symmetry energy.  

\begin{figure}
\centering
\includegraphics[width=\columnwidth]{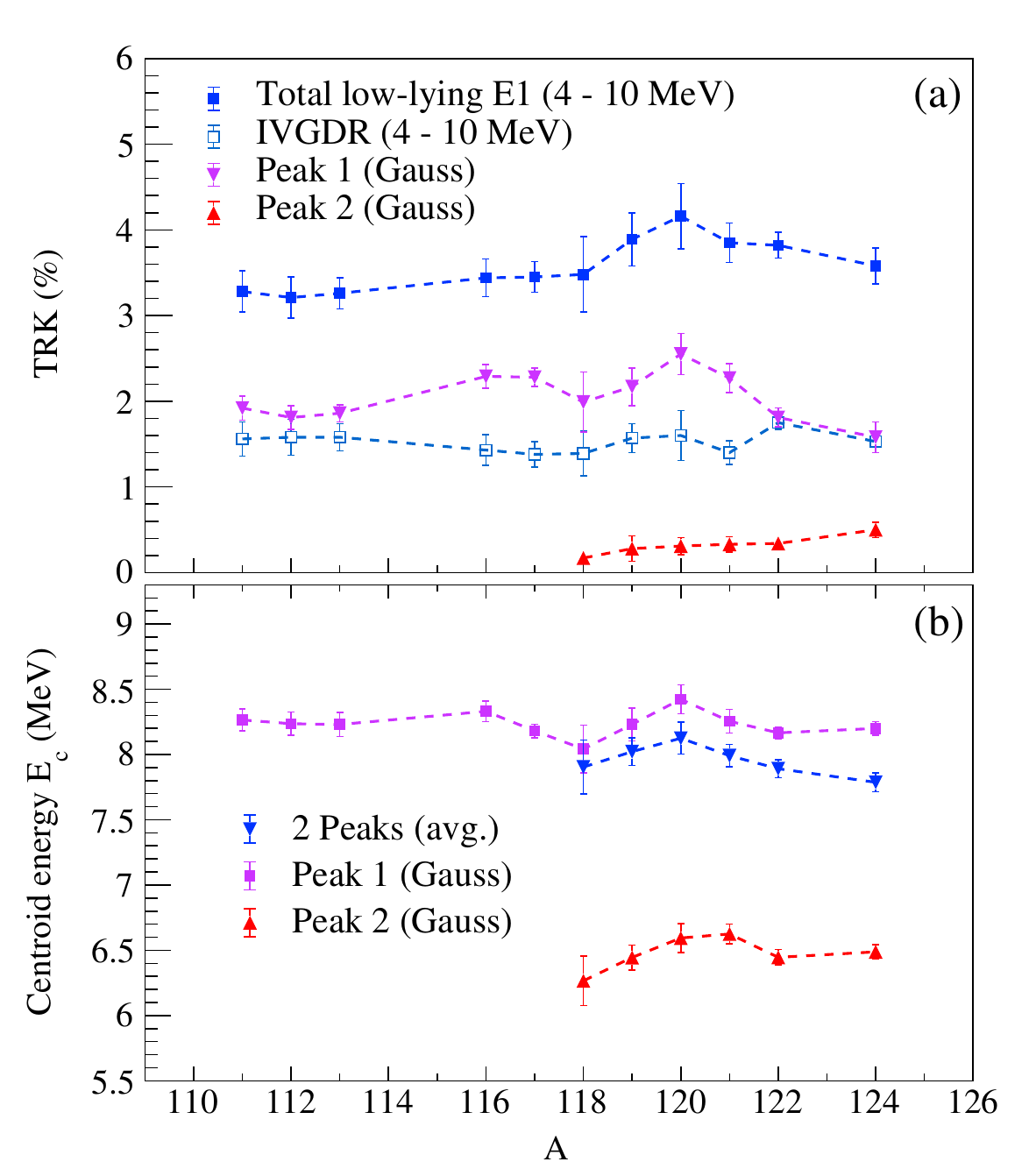}
\caption{Systematics of the isovector low-energy $E1$ strength integrated over the energy region $4 - 10$ MeV and its decomposition into the contributions from the tail of the isovector GDR and one or two (for masses $ \geq 118$) resonances on top. 
(a) Strengths in \% of the TRK sum rule. 
(b) Centroid energies.
Reprinted from Ref.~\cite{markova25}.
}
\label{fig:9}
\end{figure}

The Sn isotope chain, where the proton shell closure stabilizes the low-energy structure, is probably the best case to investigate the role of neutron excess in the PDR strength.
Results from a recent study covering the masses $A = 111 - 124$ sheds new light on this problem \cite{markova25}.
Combining studies of the low-energy isovector $E1$ strength from relativistic Coulomb excitation \cite{bassauer20a} and from the Oslo method \cite{markova24} permit a decomposition into the low-energy tail of the GDR and additional resonance-like structures.
Integrating over an energy region $4 -10$ MeV, the resulting strengths in terms of the TRK sum rule and centroid energies are displayed in Figs.~\ref{fig:9}(a) and (b), respectively.
Besides the approximately constant contribution from the GDR, a pronounced resonance around 8 MeV with a strength of about 2\% of the TRK sum rule is consistently observed. 
For $A \geq 118$ a second resonance peaked at about 6.5 MeV with much smaller strength ($\leq 0.5$\%) is revealed by the data.

The latter peak can be identified as the PDR based on the following experimental signatures:
PDR states dominantly decay to the ground state and a peak-like structure around 6.5 MeV is observed in $(\gamma,\gamma^\prime)$ experiments on Sn isotopes \cite{oezeltashenov14}.
A key feature of the PDR is its simultaneous observation in experiments with isovector and isoscalar probes.
This was demonstrated for the energy region around 6.5 MeV while the peak at higher excitation energies is absent in the isoscalar response \cite{endres10,pellegri14},  
The latter also has a different structure with dominant statistical decay \cite{krumbholz15,muescher20}.

%QRPA calculations grossly overestimate the isovector PDR strength.
%They are rather a measure of the total low-energy response.
QRPA calculations predict a correlation between the low-energy IV strength and the symmetry energy parameters \cite{carbone10}, and interactions with reasonable values compared to experimental constraints \cite{lattimer23} describe the data in the Sn isotopic chain fairly well \cite{paar25}.
However, the theoretical strength is typically concentrated in a single peak while the data show broad distributions.
In order to achieve a  more realistic description one has to invoke the coupling to complex configurations \cite{oezeltashenov14,markova25}.
This also changes the predicted transition densities in such a way that in the PDR region the characteristic "PDR-like" pattern described above is found while at higher excitation energies they change to isovector dominance \cite{markova25}.

The small isovector strengths associated with the PDR in Ref.~\cite{markova25} cast doubt on the picture of a collective neutron-skin oscillation and a possible connection to bulk properties like the symmetry energy \cite{carbone10}.
However, isoscalar cross sections are large when compared to those exciting the isoscalar GDR in the same reaction \cite{poelhekken92}.
It is also theoretically predicted that a possible collectivity of the PDR rather manifests in the isoscalar channel \cite{lanza23}.
Thus, it would be worthwhile to measure the low-energy isoscalar dipole strength along the Sn isotopic chain.

An alternative interpretation describes the PDR as a low-energy branch of the toroidal $E1$ mode \cite{repko13}.
Such a mode is consistently predicted in fluid-dynamical \cite{bastrukov93} and microscopic density functional  \cite{vretenar02,ryezayeva02} models and should be a general phenomenon across the nuclear chart \cite{repko19}.
Experimental evidence of the toroidal mode was lacking so far because it is very fragmented and strongly mixed with the isoscalar and isovector GDR. 
However, low-energy toroidal strength is predicted to be relatively pure \cite{nesterenko18}.

Recently, first experimental evidence for low-energy toroidal $E1$ transitions has been presented in the nucleus $^{58}$Ni based on a combined analysis of high-resolution $(\gamma,\gamma^\prime)$ \cite{shizuma24}, $(e,e^\prime)$ \cite{mettner87} and $(p,p^\prime)$ \cite{brandherm24} experiments.
All of these reactions are selective to dipole excitations in the proper kinematics, but methods to distinguish $E1$ and $M1$ transitions must be employed. 
In photon scattering it can be achieved by the use of polarized beams \cite{zilges22} while in proton scattering the distinction is based on a multipole decomposition analysis \cite{vonneumanncosel19}. 
In electron scattering dominance of $M1$ transitions is expected under backward angles, where the cross section contributions due to the  transverse form factor are large.

\begin{figure}
\centering
\includegraphics[width=\columnwidth]{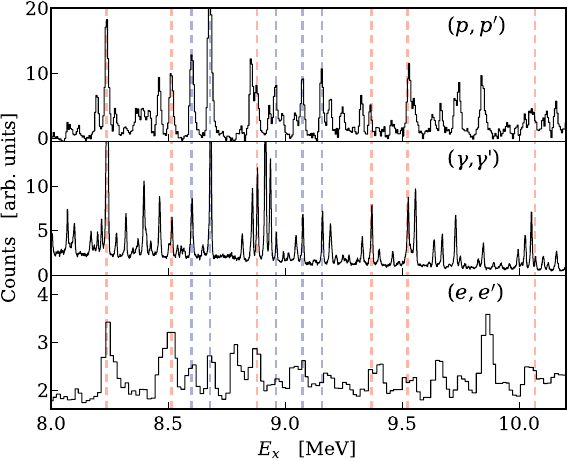}
\caption{Comparison of spectra from the $^{58}$Ni($p,p^\prime$), ($\gamma,\gamma^\prime$) and ($e,e^\prime$) reactions in the excitation energy range $8–10$ MeV for kinematics where dipole states are selectively excited. 
The vertical lines indicate dipole transitions seen in all three experiments. 
Their $E1$ (red) or $M1$ (blue) multipolarity is based on the combined analysis of the ($p,p^\prime$) and ($\gamma,\gamma^\prime$) data.
Reprinted from Ref.~\cite{brandherm24}.
}
\label{fig:10}
\end{figure}

Figure \ref{fig:10} shows representative spectra from the three reactions for excitation energies $8 - 10$ MeV and the multipolarity of dipole transitions commonly observed \cite{brandherm24}. 
The red and blue vertical dashed lines indicate $E1$ and $M1$ character based on the combined information from photon and proton scattering, respectively.
While the results for $M1$ transitions agree fairly well \cite{brandherm24}, it is evident that some of the prominent transitions seen in $(e,e^\prime)$ have electric character contrary to the assumptions in the original publication \cite{mettner87}.

\begin{figure}
\centering
\includegraphics[width=\columnwidth]{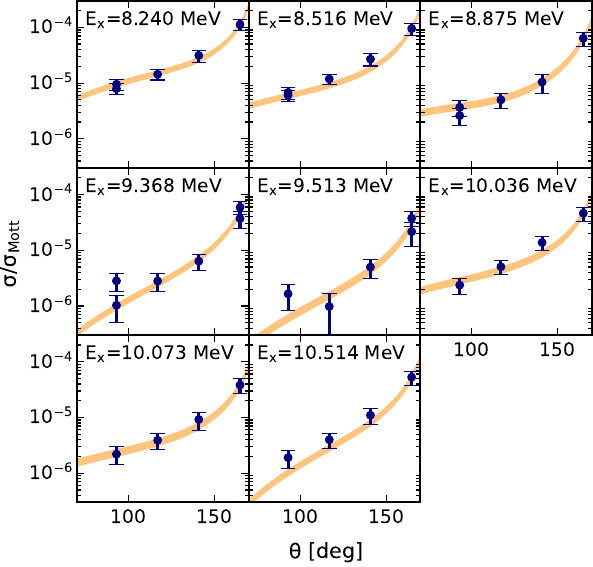}
\caption{Electron scattering cross sections (blue circles) of all
toroidal candidates identified in Ref.~\cite{brandherm24} compared with the best results from $\chi^2$ fits of all theoretical $E1$ excitations from the model described in Ref.~\cite{vonneumanncosel24} below 11 MeV (orange
bands).
Reprinted from Ref.~\cite{vonneumanncosel24}.}
\label{fig:11}
\end{figure}

These transitions are candidates for the toroidal $E1$ mode as illustrated in Fig.~\ref{fig:11}.
The measured electron scattering cross sections all show a strong increase towards backward angles indicating large transverse form factors.
Transverse form factors are a measure of the  transition current distributions.
These are expected to be negligible for the isocalar and isovector GDR which represent lateral (i.e., irrotational) motion.
On the other hand, rotational current distributions are characteristic for toroidal motion.
Theoretical predictions of the electron scattering cross sections with density functional theory can describe the experimental data in $^{58}$Ni very well and suggest that transverse $(e,e^\prime)$ form factors are a robust signature of the toroidal $E1$ mode with limited sensitivity to the chosen interaction (for details see Ref.~\cite{vonneumanncosel24}).

How does a study of $^{58}$Ni with $N \approx Z$ relate to the discussion of the PDR in nuclei with sufficient neutron excess?
The QRPA models successfully describing the toroidal
mode in $^{58}$Ni \cite{vonneumanncosel24} also predict a toroidal nature of the PDR in neutron-rich nuclei \cite{repko19}. 
Experimentally, the toroidal states identified in $^{58}$Ni exhibit the same features as low-lying $1^-$ states in neutron-rich nuclei forming the PDR, namely large
ground-state branching ratios \cite{brandherm24}, large $B(E1)$ strengths with respect to average low-energy isovector strengths and a strong isoscalar response \cite{poelhekken92}. 
Since the toroidal $E1$ mode is generic, these findings also suggest a toroidal nature of PDR modes.

A measurement of transverse electron scattering form factors in a nucleus with large neutron excess would be an important step to resolve this long-standing issue.
A recently commissioned setup for $(e,e^\prime\gamma)$ experiments \cite{hesbacher24} at the S-DALINAC facility \cite{pietralla18} promises such results and opens access to additional independent signatures of toroidal $E1$ transitions like the sign of the interference term between longitudinal and transverse form factors \cite{vonneumanncosel24}.

\section{Summary}
Angela Bracco's work in nuclear structure physics spans a wide range of topics, but a recurrent theme throughout her scientific career was the study of the nuclear electric dipole response, understanding giant and pygmy contributions, at high and low excitation energies, in hot and cold nuclei, with isoscalar and isovector probes.
Illustrated by three examples, I tried to demonstrate how her work has impacted on current topics in nuclear structure: the damping of the GDR, the equivalence of photonuclear absorption and emission at zero and finite temperatures, expressed as generalized Brink-Axel hypothesis, and the structure underlying the PDR phenomenon.   
None of these topics are solved, but her work has made esssential contributions and helped to shape the progress towards their understanding. 

\section*{Acknowledgements}
I thank the nuclear physics group at the University of Oslo for their kind hospitality during a stay where the present manuscript was prepared.
This work was supported by the Deutsche Forschungsgemeinschaft (DFG, German Research Foundation) under Contract No.\ SFB 1245 (Project ID No.\ 79384907) and by the Research Council of Norway through its grant to the Norwegian Nuclear Research Centre (Project No.\ 341985).

\bibliography{AB-bibliography}

\begin{thebibliography}{10}
\providecommand{\url}[1]{{#1}}
\providecommand{\urlprefix}{URL }
\providecommand{\doi}[1]{\url{https://doi.org/#1}}


\bibitem{bracco83}
A.~Bracco, et~al., Study of two-nucleon wave functions in $^{3}\mathrm{He}$.
\newblock Phys. Rev. Lett. \textbf{50}, 1741 (1983).
\newblock \doi{10.1103/PhysRevLett.50.1741}

\bibitem{bracco84}
A.~Bracco, et~al., Comparison of the $^3${H}e$(p,2p)d$ and $^3${H}e$(p,pd)p$
  reactions.
\newblock Phys. Lett. B \textbf{137}, 311 (1984).
\newblock \doi{10.1016/0370-2693(84)91722-2}

\bibitem{leoni17}
S.~Leoni, et~al., Multifaceted quadruplet of low-lying spin-zero states in
  $^{66}\mathrm{Ni}$: Emergence of shape isomerism in light nuclei.
\newblock Phys. Rev. Lett. \textbf{118}, 162502 (2017).
\newblock \doi{10.1103/PhysRevLett.118.162502}

\bibitem{lagergren01}
K.~Lagergren, et~al., Coexistence of superdeformed shapes in $^{154}${E}r.
\newblock Phys. Rev. Lett. \textbf{87}, 022502 (2001).
\newblock \doi{10.1103/PhysRevLett.87.022502}

\bibitem{jensen02}
D.R. Jensen, et~al., Evidence for second-phonon nuclear wobbling.
\newblock Phys. Rev. Lett. \textbf{89}, 142503 (2002).
\newblock \doi{10.1103/PhysRevLett.89.142503}

\bibitem{zanon23}
I.~Zanon, et~al., High-precision spectroscopy of $^{20}\mathrm{O}$ benchmarking
  ab initio calculations in light nuclei.
\newblock Phys. Rev. Lett. \textbf{131}, 262501 (2023).
\newblock \doi{10.1103/PhysRevLett.131.262501}

\bibitem{paxman25}
C.J. Paxman, et~al., Probing exotic cross-shell interactions at ${N}=28$ with
  single-neutron transfer on $^{47}\mathrm{K}$.
\newblock Phys. Rev. Lett. \textbf{134}, 162504 (2025).
\newblock \doi{10.1103/PhysRevLett.134.162504}

\bibitem{bracco88}
A.~Bracco, J.R. Beene, N.~Van~Giai, P.F. Bortignon, F.~Zardi, R.A. Broglia,
  Study of the breathing mode of $^{208}\mathrm{Pb}$ through neutron decay.
\newblock Phys. Rev. Lett. \textbf{60}, 2603 (1988).
\newblock \doi{10.1103/PhysRevLett.60.2603}

\bibitem{bracco89}
A.~Bracco, et~al., Saturation of the width of the giant dipole resonance at
  high temperature.
\newblock Phys. Rev. Lett. \textbf{62}, 2080 (1989).
\newblock \doi{10.1103/PhysRevLett.62.2080}

\bibitem{bortignon91}
P.F. Bortignon, A.~Bracco, D.~Brink, R.A. Broglia, Limiting temperature for the
  existence of collective motion in hot nuclei.
\newblock Phys. Rev. Lett. \textbf{67}, 3360 (1991).
\newblock \doi{10.1103/PhysRevLett.67.3360}

\bibitem{bracco02}
A.~Bracco, S.~Leoni, High-lying collective rotational states in nuclei.
\newblock Rep. Prog. Phys. \textbf{65}, 299 (2002).
\newblock \doi{10.1088/0034-4885/65/2/204}

\bibitem{bracco21}
A.~Bracco, G.~Duchêne, Z.~Podolyák, P.~Reiter, Gamma spectroscopy with
  {AGATA} in its first phases: {N}ew insights in nuclear excitations along the
  nuclear chart.
\newblock Prog. Part. Nucl. Phys. \textbf{121}, 103887 (2021).
\newblock \doi{https://doi.org/10.1016/j.ppnp.2021.103887}

\bibitem{bracco19}
A.~Bracco, E.~Lanza, A.~Tamii, Isoscalar and isovector dipole excitations:
  Nuclear properties from low-lying states and from the isovector giant dipole
  resonance.
\newblock Prog. Part. Nucl. Phys. \textbf{106}, 360 (2019).
\newblock \doi{10.1016/j.ppnp.2019.02.001}

\bibitem{harakeh01}
M.N. Harakeh, A.~Van~der Woude, \emph{Giant {R}esonances: {F}undamental
  {H}igh-{F}requency {M}odes of {N}uclear {E}xcitation} (Oxford Univ. Press,
  Oxford, 2001)

\bibitem{vonneumanncosel19}
P.~von Neumann-Cosel, A.~Tamii, Electric and magnetic dipole modes in
  high-resolution inelastic proton scattering at 0{\textdegree}.
\newblock Eur. Phys. J. A \textbf{55}, 110 (2019).
\newblock \doi{10.1140/epja/i2019-12781-7}

\bibitem{beene90}
J.R. Beene, et~al., Heavy-ion {C}oulomb excitation and photon decay of the
  giant dipole resonance in $^{208}\mathrm{Pb}$.
\newblock Phys. Rev. C \textbf{41}, 920 (1990).
\newblock \doi{10.1103/PhysRevC.41.920}

\bibitem{bolme88}
G.O. Bolme, et~al., ($e,e^\prime n$) coincidence studies of the giant multipole
  resonances of $^{208}\mathrm{Pb}$.
\newblock Phys. Rev. Lett. \textbf{61}, 1081 (1988).
\newblock \doi{10.1103/PhysRevLett.61.1081}

\bibitem{diesener94}
H.~Diesener, et~al., Excitation and decay of giant resonances in the
  $^{40}\mathrm{Ca}$$(e,e'x)$ reaction.
\newblock Phys. Rev. Lett. \textbf{72}, 1994 (1994).
\newblock \doi{10.1103/PhysRevLett.72.1994}

\bibitem{strauch00}
S.~Strauch, et~al., Giant resonances in the doubly magic nucleus $^{48}${C}a
  from the $(e,e^\prime n)$ reaction.
\newblock Phys. Rev. Lett. \textbf{85}, 2913 (2000).
\newblock \doi{10.1103/PhysRevLett.85.2913}

\bibitem{hunyadi03}
M.~Hunyadi, et~al., Direct proton decay of the isoscalar giant dipole
  resonance.
\newblock Phys. Lett. B \textbf{576}, 253 (2003).
\newblock \doi{10.1016/j.physletb.2003.10.016}

\bibitem{bortignon98}
P.~Bortignon, A.~Bracco, R.~Broglia, \emph{Giant Resonances: {N}uclear
  Structure at finite Temperature} (CRC Press, London, 1998)

\bibitem{vonneumanncosel19a}
P.~von Neumann-Cosel, V.Y. Ponomarev, A.~Richter, J.~Wambach, Gross,
  intermediate and fine structure of nuclear giant resonances: Evidence for
  doorway states.
\newblock Eur. Phys. J. A \textbf{55}, 224 (2019).
\newblock \doi{10.1140/epja/i2019-12795-1}

\bibitem{olorunfunmi22}
S.D. Olorunfunmi, et~al., Evolution of the isoscalar giant monopole resonance
  in the {C}a isotope chain.
\newblock Phys. Rev. C \textbf{105}, 054319 (2022).
\newblock \doi{10.1103/PhysRevC.105.054319}

\bibitem{bahini24}
A.~Bahini, et~al., Fine structure of the isoscalar giant monopole resonance in
  $^{58}\mathrm{Ni}, ^{90}\mathrm{Zr}, ^{120}\mathrm{Sn}$, and
  $^{208}\mathrm{Pb}$.
\newblock Phys. Rev. C \textbf{109}, 014325 (2024).
\newblock \doi{10.1103/PhysRevC.109.014325}

\bibitem{poltoratska14}
I.~Poltoratska, et~al., Fine structure of the isovector giant dipole resonance
  in $^{208}\mathrm{Pb}$: Characteristic scales and level densities.
\newblock Phys. Rev. C \textbf{89}, 054322 (2014).
\newblock \doi{10.1103/PhysRevC.89.054322}

\bibitem{jingo18}
M.~Jingo, et~al., Studies of the giant dipole resonance in $^{27}${A}l ,
  $^{40}${C}a , $^{56}${F}e , $^{58}${N}i and $^{208}${P}b with high
  energy-resolution inelastic proton scattering under 0{\textdegree}.
\newblock Eur. Phys. J. A \textbf{54}, 234 (2018).
\newblock \doi{10.1140/epja/i2018-12664-5}

\bibitem{donaldson21}
L.M. Donaldson, et~al., Fine structure of the isovector giant dipole resonance
  in $^{142\text{--}150}\mathrm{Nd}$ and $^{152}\mathrm{Sm}$.
\newblock Phys. Rev. C \textbf{102}, 064327 (2020).
\newblock \doi{10.1103/PhysRevC.102.064327}

\bibitem{shevchenko04}
A.~Shevchenko, et~al., Fine structure in the energy region of the isoscalar
  giant quadrupole resonance: Characteristic scales from a wavelet analysis.
\newblock Phys. Rev. Lett. \textbf{93}, 122501 (2004).
\newblock \doi{10.1103/PhysRevLett.93.122501}

\bibitem{shevchenko09}
A.~Shevchenko, et~al., Global investigation of the fine structure of the
  isoscalar giant quadrupole resonance.
\newblock Phys. Rev. C \textbf{79}, 044305 (2009).
\newblock \doi{10.1103/PhysRevC.79.044305}

\bibitem{kalmykov06}
Y.~Kalmykov, et~al., Fine structure of the {G}amow-{T}eller resonance in
  $^{90}\mathrm{Nb}$ and level density of ${1}^{+}$ states.
\newblock Phys. Rev. Lett. \textbf{96}, 012502 (2006).
\newblock \doi{10.1103/PhysRevLett.96.012502}

\bibitem{vonneumanncosel99}
P.~von Neumann-Cosel, et~al., Spin and orbital magnetic quadrupole resonances
  in $^{48}${Ca} and $^{90}${Z}r from $180\ifmmode^\circ\else\textdegree\fi{}$
  electron scattering.
\newblock Phys. Rev. Lett. \textbf{82}, 1105 (1999).
\newblock \doi{10.1103/PhysRevLett.82.1105}

\bibitem{kureba18}
C.~Kureba, et~al., Wavelet signatures of ${K}$-splitting of the isoscalar giant
  quadrupole resonance in deformed nuclei from high-resolution $(p,p^\prime)$
  scattering off $^{146,148,150}${N}d.
\newblock Phys. Lett. B \textbf{779}, 269 (2018).
\newblock \doi{10.1016/j.physletb.2018.02.013}

\bibitem{bahini23}
A.~Bahini, et~al., Isoscalar giant monopole strength in $^{58}\mathrm{Ni},
  ^{90}\mathrm{Zr}, ^{120}\mathrm{Sn}$, and $^{208}\mathrm{Pb}$.
\newblock Phys. Rev. C \textbf{107}, 034312 (2023).
\newblock \doi{10.1103/PhysRevC.107.034312}

\bibitem{tamii11}
A.~Tamii, et~al., Complete electric dipole response and the neutron skin in
  $^{208}\mathrm{Pb}$.
\newblock Phys. Rev. Lett. \textbf{107}, 062502 (2011).
\newblock \doi{10.1103/PhysRevLett.107.062502}

\bibitem{aiba99}
H.~Aiba, M.~Matsuo, Scaling analysis of the fluctuating strength function.
\newblock Phys. Rev. C \textbf{60}, 034307 (1999).
\newblock \doi{10.1103/PhysRevC.60.034307}

\bibitem{aiba11}
H.~Aiba, M.~Matsuo, S.~Nishizaki, T.~Suzuki, Fluctuation properties of the
  strength function associated with the giant quadrupole resonance in
  $^{208}\mathrm{Pb}$.
\newblock Phys. Rev. C \textbf{83}, 024314 (2011).
\newblock \doi{10.1103/PhysRevC.83.024314}

\bibitem{lacroix99}
D.~Lacroix, P.~Chomaz, Multiscale fluctuations in the nuclear response.
\newblock Phys. Rev. C \textbf{60}, 064307 (1999).
\newblock \doi{10.1103/PhysRevC.60.064307}

\bibitem{lacroix00}
D.~Lacroix, A.~Mai, P.~{von Neumann-Cosel}, A.~Richter, J.~Wambach, Multiple
  scales in the fine structure of the isoscalar giant quadrupole resonance in
  $^{208}${P}b.
\newblock Phys. Lett. B \textbf{479}, 15 (2000).
\newblock \doi{10.1016/S0370-2693(00)00301-4}

\bibitem{heiss11}
W.D. Heiss, R.G. Nazmitdinov, F.D. Smit, Time scales in nuclear giant
  resonances.
\newblock Phys. Rev. C \textbf{81}, 034604 (2010).
\newblock \doi{10.1103/PhysRevC.81.034604}

\bibitem{shevchenko08}
A.~Shevchenko, et~al., Analysis of fine structure in the nuclear continuum.
\newblock Phys. Rev. C \textbf{77}, 024302 (2008).
\newblock \doi{10.1103/PhysRevC.77.024302}

\bibitem{carter22}
J.~Carter, et~al., Damping of the isovector giant dipole resonance in
  $^{40,48}${C}a.
\newblock Phys. Lett. B \textbf{833}, 137374 (2022).
\newblock \doi{10.1016/j.physletb.2022.137374}

\bibitem{usman11}
I.~Usman, et~al., Fine structure of the isoscalar giant quadrupole resonance in
  $^{40}${C}a due to {L}andau damping?
\newblock Phys. Lett. B \textbf{698}, 191 (2011).
\newblock \doi{10.1016/j.physletb.2011.03.015}

\bibitem{johnson15}
C.W. Johnson, Systematics of strength function sum rules.
\newblock Phys. Lett. B \textbf{750}, 72 (2015).
\newblock \doi{10.1016/j.physletb.2015.08.054}

\bibitem{sieja23}
K.~Sieja, {B}rink-{A}xel hypothesis in the pygmy-dipole resonance region.
\newblock Eur. Phys. J. A \textbf{59}, 147 (2023).
\newblock \doi{10.1140/epja/s10050-023-01067-8}

\bibitem{hung17}
N.Q. Hung, N.D. Dang, L.T.Q. Huong, Simultaneous microscopic description of
  nuclear level density and radiative strength function.
\newblock Phys. Rev. Lett. \textbf{118}, 022502 (2017).
\newblock \doi{10.1103/PhysRevLett.118.022502}

\bibitem{wibovo22}
H.~Wibowo, E.~Litvinova, Nuclear shell structure in a finite-temperature
  relativistic framework.
\newblock Phys. Rev. C \textbf{106}, 044304 (2022).
\newblock \doi{10.1103/PhysRevC.106.044304}

\bibitem{kaur25}
A.~Kaur, E.~Y\"uksel, N.~Paar, Electric and magnetic $\ensuremath{\gamma}$-ray
  strength functions at finite temperature.
\newblock Phys. Rev. C \textbf{112}, 014307 (2025).
\newblock \doi{10.1103/96g9-1ff5}

\bibitem{markova22}
M.~Markova, et~al., Nuclear level densities and $\ensuremath{\gamma}$-ray
  strength functions in $^{120,124}\mathrm{Sn}$ isotopes: {I}mpact of
  {P}orter-{T}homas fluctuations.
\newblock Phys. Rev. C \textbf{106}, 034322 (2022).
\newblock \doi{10.1103/PhysRevC.106.034322}

\bibitem{brink55}
D.M. Brink.
\newblock Some aspects of the interaction of fields with matter.
\newblock Doctoral thesis, Oxford University (1955).

\bibitem{axel62}
P.~Axel, Electric dipole ground-state transition width strength function and
  7-{M}e{V} photon interactions.
\newblock Phys. Rev. \textbf{126}, 671 (1962).
\newblock \doi{10.1103/PhysRev.126.671}

\bibitem{broglia92}
R.~Broglia, P.~Bortignon, A.~Bracco, The giant dipole resonance in hot nuclei.
\newblock Prog. Part. Nucl. Phys. \textbf{28}, 517 (1992).
\newblock \doi{10.1016/0146-6410(92)90054-6}

\bibitem{bracco95}
A.~Bracco, et~al., Increase in width of the giant dipole resonance in hot
  nuclei: {S}hape change or collisional damping?
\newblock Phys. Rev. Lett. \textbf{74}, 3748 (1995).
\newblock \doi{10.1103/PhysRevLett.74.3748}

\bibitem{guttormsen16}
M.~Guttormsen, et~al., Validity of the generalized {B}rink-{A}xel hypothesis in
  $^{238}\mathrm{Np}$.
\newblock Phys. Rev. Lett. \textbf{116}, 012502 (2016).
\newblock \doi{10.1103/PhysRevLett.116.012502}

\bibitem{martin17}
D.~Martin, et~al., Test of the {B}rink-{A}xel hypothesis for the pygmy dipole
  resonance.
\newblock Phys. Rev. Lett. \textbf{119}, 182503 (2017).
\newblock \doi{10.1103/PhysRevLett.119.182503}

\bibitem{crespocampo18}
L.~Crespo~Campo, et~al., Test of the generalized {B}rink-{A}xel hypothesis in
  $^{64,65}\mathrm{Ni}$.
\newblock Phys. Rev. C \textbf{98}, 054303 (2018).
\newblock \doi{10.1103/PhysRevC.98.054303}

\bibitem{scholz20}
P.~Scholz, et~al., Primary $\ensuremath{\gamma}$-ray intensities and
  $\ensuremath{\gamma}$-strength functions from discrete two-step
  $\ensuremath{\gamma}$-ray cascades in radiative proton-capture experiments.
\newblock Phys. Rev. C \textbf{101}, 045806 (2020).
\newblock \doi{10.1103/PhysRevC.101.045806}

\bibitem{angell12}
C.T. Angell, et~al., Evidence for radiative coupling of the pygmy dipole
  resonance to excited states.
\newblock Phys. Rev. C \textbf{86}, 051302 (2012).
\newblock \doi{10.1103/PhysRevC.86.051302}

\bibitem{isaak13}
J.~Isaak, et~al., Constraining nuclear photon strength functions by the decay
  properties of photo-excited states.
\newblock Phys.Lett. B \textbf{727}(4), 361 (2013).
\newblock \doi{10.1016/j.physletb.2013.10.040}

\bibitem{netterdon15}
L.~Netterdon, et~al., Experimental constraints on the $\gamma$-ray strength
  function in $^{90}${Z}r using partial cross sections of the
  $^{89}${Y}(p,$\gamma$)$^{90}$zr reaction.
\newblock Phys. Lett. B \textbf{744}, 358 (2015).
\newblock \doi{10.1016/j.physletb.2015.04.018}

\bibitem{isaak19}
J.~Isaak, et~al., The concept of nuclear photon strength functions: A
  model-independent approach via ($\vec{\gamma}, \gamma^\prime
  \gamma^{\prime\prime}$) reactions.
\newblock Phys. Lett. B \textbf{788}, 225 (2019).
\newblock \doi{10.1016/j.physletb.2018.11.038}

\bibitem{bassauer20a}
S.~Bassauer, et~al., Electric and magnetic dipole strength in
  $^{112,114,116,118,120,124}\mathrm{Sn}$.
\newblock Phys. Rev. C \textbf{102}, 034327 (2020).
\newblock \doi{10.1103/PhysRevC.102.034327}

\bibitem{markova21}
M.~Markova, et~al., Comprehensive test of the {B}rink-{A}xel hypothesis in the
  energy region of the pygmy dipole resonance.
\newblock Phys. Rev. Lett. \textbf{127}, 182501 (2021).
\newblock \doi{10.1103/PhysRevLett.127.182501}

\bibitem{schiller00}
A.~Schiller, L.~Bergholt, M.~Guttormsen, E.~Melby, J.~Rekstad, S.~Siem,
  Extraction of level density and $\gamma$ strength function from primary
  $\gamma$ spectra.
\newblock Nucl. Instrum. Methods A \textbf{447}, 498 (2000).
\newblock \doi{10.1016/S0168-9002(99)01187-0}

\bibitem{wiedeking21}
M.~Wiedeking, et~al., Independent normalization for $\ensuremath{\gamma}$-ray
  strength functions: The shape method.
\newblock Phys. Rev. C \textbf{104}, 014311 (2021).
\newblock \doi{10.1103/PhysRevC.104.014311}

\bibitem{savran13}
D.~Savran, T.~Aumann, A.~Zilges, Experimental studies of the pygmy dipole
  resonance.
\newblock Prog. Part. Nucl. Phys. \textbf{70}, 210 (2013).
\newblock \doi{10.1016/j.ppnp.2013.02.003}

\bibitem{wiedeking24}
M.~Wiedeking, S.~Goriely, Photon strength functions and nuclear level
  densities: {I}nvaluable input for nucleosynthesis.
\newblock Phil. Trans. R. Soc. A. \textbf{382}, 20230125 (2024).
\newblock \doi{doi/10.1098/rsta.2023.0125}

\bibitem{wieland09}
O.~Wieland, et~al., Search for the pygmy dipole resonance in $^{68}\mathrm{Ni}$
  at 600 {MeV}/nucleon.
\newblock Phys. Rev. Lett. \textbf{102}, 092502 (2009).
\newblock \doi{10.1103/PhysRevLett.102.092502}

\bibitem{wieland18}
O.~Wieland, et~al., Low-lying dipole response in the unstable
  $^{70}\mathrm{Ni}$ nucleus.
\newblock Phys. Rev. C \textbf{98}, 064313 (2018).
\newblock \doi{10.1103/PhysRevC.98.064313}

\bibitem{crespi14}
F.C.L. Crespi, et~al., Isospin character of low-lying pygmy dipole states in
  $^{208}\mathrm{Pb}$ via inelastic scattering of $^{17}\mathrm{O}$ ions.
\newblock Phys. Rev. Lett. \textbf{113}, 012501 (2014).
\newblock \doi{10.1103/PhysRevLett.113.012501}

\bibitem{pellegri14}
L.~Pellegri, et~al., Pygmy dipole resonance in $^{124}${S}n populated by
  inelastic scattering of $^{17}${O}.
\newblock Phys. Lett. B \textbf{738}, 519 (2014).
\newblock \doi{10.1016/j.physletb.2014.08.029}

\bibitem{crespi21}
F.~Crespi, et~al., The structure of low-lying $1^-$ states in $^{90,94}${Z}r
  from ($\alpha,\alpha^\prime\gamma$) and ($p,p^\prime\gamma$) reactions.
\newblock Phys. Lett. B \textbf{816}, 136210 (2021).
\newblock \doi{10.1016/j.physletb.2021.136210}

\bibitem{carbone10}
A.~Carbone, et~al., Constraints on the symmetry energy and neutron skins from
  pygmy resonances in $^{68}\mathrm{Ni}$ and $^{132}\mathrm{Sn}$.
\newblock Phys. Rev. C \textbf{81}, 041301 (2010).
\newblock \doi{10.1103/PhysRevC.81.041301}

\bibitem{lanza23}
E.~Lanza, L.~Pellegri, A.~Vitturi, M.~Andrés, Theoretical studies of pygmy
  resonances.
\newblock Prog. Part. Nucl. Phys. \textbf{129}, 104006 (2023).
\newblock \doi{10.1016/j.ppnp.2022.104006}

\bibitem{paar07}
N.~Paar, D.~Vretenar, E.~Khan, G.~Colò, Exotic modes of excitation in atomic
  nuclei far from stability.
\newblock Rep. Prog. Phys. \textbf{70}, R02 (2007).
\newblock \doi{10.1088/0034-4885/70/5/R02}

\bibitem{rocamaza18}
X.~Roca-Maza, N.~Paar, Nuclear equation of state from ground and collective
  excited state properties of nuclei.
\newblock Prog. Part. Nucl. Phys. \textbf{101}, 96 (2018).
\newblock \doi{10.1016/j.ppnp.2018.04.001}

\bibitem{repko13}
A.~Repko, P.G. Reinhard, V.O. Nesterenko, J.~Kvasil, Toroidal nature of the
  low-energy ${E}1$ mode.
\newblock Phys. Rev. C \textbf{87}, 024305 (2013).
\newblock \doi{10.1103/PhysRevC.87.024305}

\bibitem{repko19}
A.~Repko, V.O. Nesterenko, J.~Kvasil, P.G. Reinhard, Systematics of toroidal
  dipole modes in {C}a, {N}i, {Z}r, and {S}n isotopes.
\newblock Eur. Phys. J. A \textbf{55}, 242 (2019).
\newblock \doi{10.1140/epja/i2019-12770-x}

\bibitem{poelhekken92}
T.~Poelhekken, S.~Hesmondhalgh, H.~Hofmann, A.~{van der Woude}, M.~Harakeh,
  Low-energy isoscalar dipole strength in $^{40}${C}a, $^{58}${N}i, $^{90}${Z}r
  and $^{208}${P}b.
\newblock Phys. Lett. B \textbf{278}, 423 (1992).
\newblock \doi{10.1016/0370-2693(92)90579-S}

\bibitem{markova25}
M.~Markova, P.~{von Neumann-Cosel}, E.~Litvinova, Systematics of the low-energy
  electric dipole strength in the {S}n isotopic chain.
\newblock Phys. Lett. B \textbf{860}, 139216 (2025).
\newblock \doi{10.1016/j.physletb.2024.139216}

\bibitem{markova24}
M.~Markova, et~al., Systematic study of the low-lying electric dipole strength
  in {S}n isotopes and its astrophysical implications.
\newblock Phys. Rev. C \textbf{109}, 054311 (2024).
\newblock \doi{10.1103/PhysRevC.109.054311}

\bibitem{oezeltashenov14}
B.~\"Ozel-Tashenov, et~al., Low-energy dipole strength in
  $^{112,120}\mathrm{Sn}$.
\newblock Phys. Rev. C \textbf{90}, 024304 (2014).
\newblock \doi{10.1103/PhysRevC.90.024304}

\bibitem{endres10}
J.~Endres, et~al., Isospin character of the pygmy dipole resonance in
  $^{124}\mathrm{Sn}$.
\newblock Phys. Rev. Lett. \textbf{105}, 212503 (2010).
\newblock \doi{10.1103/PhysRevLett.105.212503}

\bibitem{krumbholz15}
A.~Krumbholz, et~al., Low-energy electric dipole response in $^{120}${S}n.
\newblock Phys. Lett. B \textbf{744}, 7 (2015).
\newblock \doi{10.1016/j.physletb.2015.03.023}

\bibitem{muescher20}
M.~M\"uscher, et~al., High-sensitivity investigation of low-lying dipole
  strengths in $^{120}\mathrm{Sn}$.
\newblock Phys. Rev. C \textbf{102}, 014317 (2020).
\newblock \doi{10.1103/PhysRevC.102.014317}

\bibitem{lattimer23}
J.M. Lattimer, Constraints on nuclear symmetry energy parameters.
\newblock Particles \textbf{6}, 30 (2023).
\newblock \doi{10.3390/particles6010003}

\bibitem{paar25}
N.~Paar, A.~Kaur, Properties of the pygmy dipole strength from theoretical
  perspective.
\newblock Acta Phys. Pol. B Proc. Suppl. \textbf{18}, 2--A31 (2025).
\newblock \doi{10.5506/APhysPolBSupp.18.2-A31}

\bibitem{bastrukov93}
S.~Bastrukov, S.~Misicu, A.~Sushkov, Dipole torus mode in nuclear
  fluid-dynamics.
\newblock Nucl. Phys. A \textbf{562}, 191 (1993).
\newblock \doi{10.1016/0375-9474(93)90195-4}

\bibitem{vretenar02}
D.~Vretenar, N.~Paar, P.~Ring, T.~Nik\ifmmode \check{s}\else
  \v{s}\fi{}i\ifmmode~\acute{c}\else \'{c}\fi{}, Toroidal dipole resonances in
  the relativistic random phase approximation.
\newblock Phys. Rev. C \textbf{65}, 021301 (2002).
\newblock \doi{10.1103/PhysRevC.65.021301}

\bibitem{ryezayeva02}
N.~Ryezayeva, et~al., Nature of low-energy dipole strength in nuclei: The case
  of a resonance at particle threshold in
  $^{\mathrm{208}}\mathrm{P}\mathrm{b}$.
\newblock Phys. Rev. Lett. \textbf{89}, 272502 (2002).
\newblock \doi{10.1103/PhysRevLett.89.272502}

\bibitem{nesterenko18}
V.O. Nesterenko, A.~Repko, J.~Kvasil, P.G. Reinhard, Individual low-energy
  toroidal dipole state in $^{24}\mathrm{Mg}$.
\newblock Phys. Rev. Lett. \textbf{120}, 182501 (2018).
\newblock \doi{10.1103/PhysRevLett.120.182501}

\bibitem{shizuma24}
T.~Shizuma, et~al., Parity assignment for low-lying dipole states in
  $^{58}\mathrm{Ni}$.
\newblock Phys. Rev. C \textbf{109}, 014302 (2024).
\newblock \doi{10.1103/PhysRevC.109.014302}

\bibitem{mettner87}
W.~Mettner, A.~Richter, W.~Stock, B.~Metsch, A.~{Van Hees}, Electroexcitation
  of $^{58}${N}i: A study of the fragmentation of the magnetic dipole strength.
\newblock Nucl. Phys. A \textbf{473}, 160 (1987).
\newblock \doi{10.1016/0375-9474(87)90159-X}

\bibitem{brandherm24}
I.~Brandherm, et~al., Electric and magnetic dipole strength in
  $^{58}\mathrm{Ni}$ from forward-angle proton scattering.
\newblock Phys. Rev. C \textbf{110}, 034319 (2024).
\newblock \doi{10.1103/PhysRevC.110.034319}

\bibitem{zilges22}
A.~Zilges, D.~Balabanski, J.~Isaak, N.~Pietralla, Photonuclear reactions --
  from basic research to applications.
\newblock Prog. Part. Nucl. Phys. \textbf{122}, 103903 (2022).
\newblock \doi{10.1016/j.ppnp.2021.103903}

\bibitem{vonneumanncosel24}
P.~von Neumann-Cosel, et~al., Candidate toroidal electric dipole mode in the
  spherical nucleus $^{58}\mathrm{Ni}$.
\newblock Phys. Rev. Lett. \textbf{133}, 232502 (2024).
\newblock \doi{10.1103/PhysRevLett.133.232502}

\bibitem{hesbacher24}
B.~Hesbacher, et~al., The electron-gamma coincidence setup {DAGOBERT}.
\newblock Nucl. Instrum. Methods A \textbf{1078}, 170574 (2025).
\newblock \doi{10.1016/j.nima.2025.170574}

\bibitem{pietralla18}
N.~Pietralla, The {I}nstitute of {N}uclear {P}hysics at the {TU} {D}armstadt.
\newblock Nucl. Phys. News \textbf{28}(2), 4 (2018).
\newblock \doi{10.1080/10619127.2018.1463013}

\end{thebibliography}

\end{document}